\documentclass[conference]{IEEEtran}
\IEEEoverridecommandlockouts
\usepackage{listings}
\usepackage{cite}
\usepackage{amsmath,amssymb,amsfonts}
\usepackage{algorithmic}
\usepackage{graphicx}
\usepackage{textcomp}
\usepackage{xcolor}

\def\BibTeX{{\rm B\kern-.05em{\sc i\kern-.025em b}\kern-.08em
    T\kern-.1667em\lower.7ex\hbox{E}\kern-.125emX}}
\begin{document}

\title{ Data-driven approaches to estimating the habitability of exoplanets\\
}

\author{\IEEEauthorblockN{Mithil Sai Jakka}
\IEEEauthorblockA{\textit{Department of Computer Science} \\
\textit{California State University}\\
Fresno, USA \\
}
}

\maketitle

\begin{abstract}
The exploration and study of exoplanets remain at the frontier of astronomical research, challenging scientists to continuously innovate and refine methodologies to navigate the vast, complex data these celestial bodies produce. This literature review aims to illuminate the emerging trends and advancements within this sphere, specifically focusing on the interplay between exoplanet detection, classification, and visualization, and the increasingly pivotal role of machine learning and computational models. Our journey through this realm of exploration commences with a comprehensive analysis of fifteen meticulously selected, seminal papers in the field. These papers, each representing a distinct facet of exoplanet research, collectively offer a multi-dimensional perspective on the current state of the field. They provide valuable insights into the innovative application of machine learning techniques to overcome the challenges posed by the analysis and interpretation of astronomical data. From the application of Support Vector Machines (SVM) to Deep Learning models, the review encapsulates the broad spectrum of machine learning approaches employed in exoplanet research. The review also seeks to unravel the story woven by the data within these papers, detailing the triumphs and tribulations of the field. It highlights the increasing reliance on diverse datasets, such as Kepler and TESS, and the push for improved accuracy in exoplanet detection and classification models. The narrative concludes with key takeaways and insights, drawing together the threads of research to present a cohesive picture of the direction in which the field is moving. This literature review, therefore, serves not just as an academic exploration, but also as a narrative of scientific discovery and innovation in the quest to understand our cosmic neighborhood.

\end{abstract}


\section{Introduction}
Astronomy, as an age-old science, has long captivated humanity with its myriad mysteries and grandeur. In recent years, one particular area of this expansive field has seized the spotlight: the exploration of exoplanets \cite{b1}. Exoplanets, celestial bodies orbiting stars beyond our own solar system, offer tantalizing prospects in the quest to understand the universe's complexities \cite{b4}. The burgeoning interest in these distant worlds reflects our insatiable curiosity about the cosmos and the fundamental question of whether life exists beyond Earth \cite{b6}.

Our understanding of these celestial bodies has been propelled by the advent of increasingly sophisticated space missions and telescopes, such as Kepler and the Transiting Exoplanet Survey Satellite (TESS) \cite{b7}. These instruments have flooded the scientific community with vast volumes of data, birthing a new era of discovery \cite{b9}. However, this abundance of data has also presented a challenge: the need for efficient, accurate methods of analysis to extract meaningful insights from the deluge.

In response to this, the science of astrophysics has intertwined with the art of data science, resulting in remarkable synergy. Computational models and machine learning techniques have emerged as powerful tools in the analysis and interpretation of astronomical data. From the detection of exoplanets to their classification, these techniques have revolutionized the field, transforming raw data into valuable knowledge \cite{b15}.

Below is just an imaginary, prediction of an exoplanet's habitability detection using the transit of the planet around its star.
\begin{figure}[htbp]
\begin{center}
\includegraphics[width=0.5\textwidth]{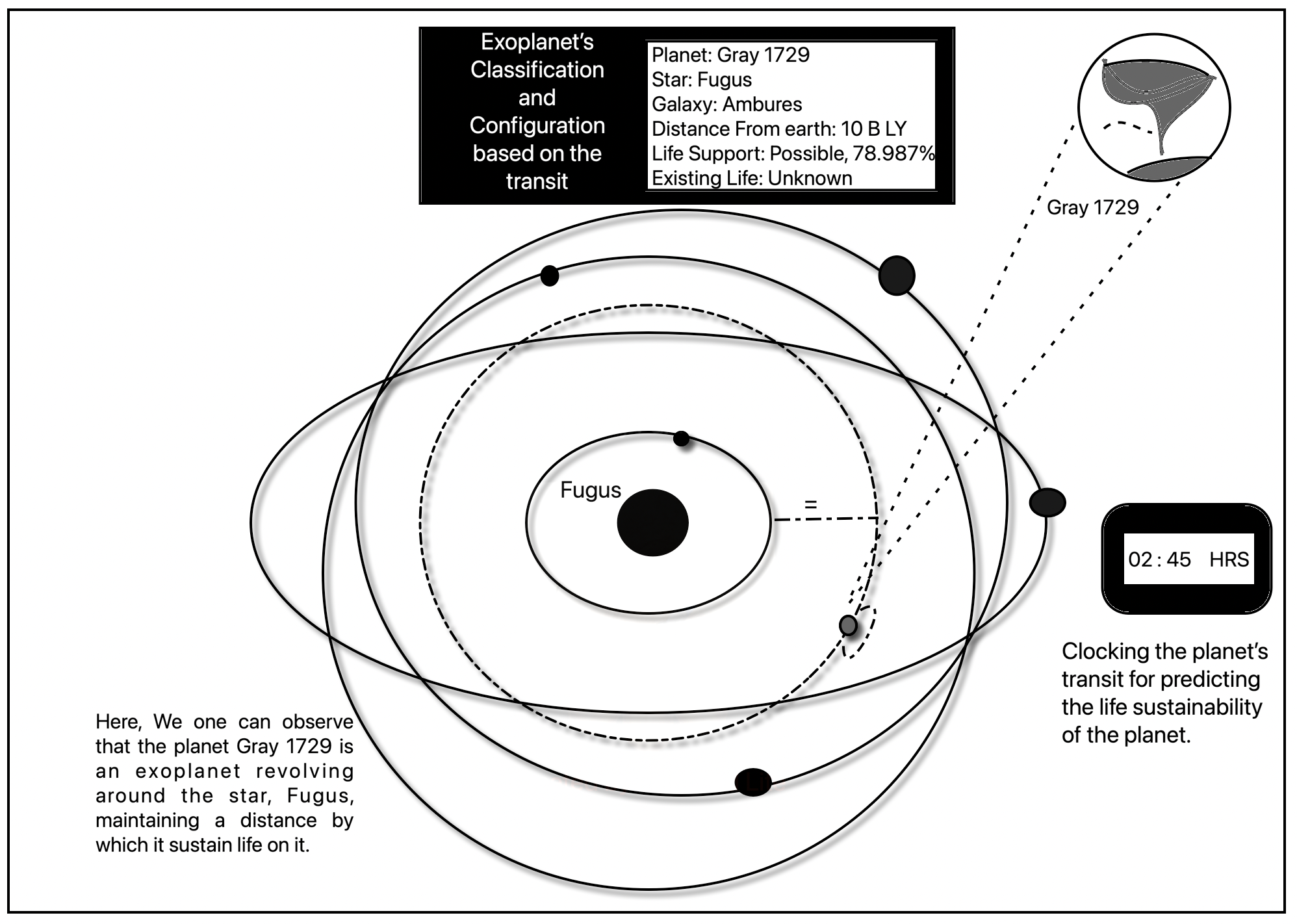}
\end{center}
\caption{Calculation of the transit of Exoplanet}
\label{fig1}
\end{figure}

This literature review navigates the convergence of astrophysics and data science, focusing on the current trends and advancements in exoplanet detection and classification \cite{b2}. It illuminates the journey from the collection of raw data to its transformation into meaningful insights, via the innovative application of machine learning and computational models. The objective is to weave a narrative that encapsulates the dynamic landscape of exoplanet research, elucidating the challenges, triumphs, and possibilities that lie ahead in this exciting field of study.

\section{Background Knowledge/Premise}
The detection and classification of exoplanets, or planets that exist outside our solar system, have long been the domain of astronomers and astrophysicists. These experts have dedicated significant resources to exploring the vast expanses of the universe, seeking to better understand the celestial bodies that populate the cosmos. In recent years, the field has experienced a veritable explosion of new data stemming from an array of space missions and telescopes, which have captured a wealth of information on these distant worlds \cite{b5}. While this data has been invaluable in expanding our knowledge of exoplanets, it has also introduced new challenges that require innovative techniques for analyzing and interpreting these massive, complex datasets.

The emergence of data science, particularly the field of machine learning, has offered promising solutions to these challenges \cite{b9}. Machine learning, a subset of artificial intelligence, involves the development of algorithms that enable computers to learn from and make predictions or decisions based on data \cite{b11}. By applying these algorithms to the study of exoplanets, researchers can effectively analyze the vast troves of information gathered by space missions and telescopes [12]. This approach allows scientists to identify patterns and correlations within the data, making it possible to detect and classify new exoplanets with greater accuracy and efficiency.

Several machine learning algorithms, such as Support Vector Machine (SVM), k-Nearest Neighbors (kNN), and K-means clustering, have proven particularly effective in the identification and categorization of exoplanets [15]. These techniques offer unique advantages in the analysis of large datasets, as they can identify patterns and relationships within the data that may not be immediately apparent through traditional methods [1], [2]. For example, SVM is a supervised learning method that can be used for classification or regression, and it excels at separating data points into distinct categories based on their features [3], [4]. Similarly, kNN is a simple yet powerful algorithm that can be used for classification or regression tasks, relying on the proximity of data points to make predictions [5], [6]. K-means clustering, on the other hand, is an unsupervised learning method that groups data points into clusters based on their similarities, which can be invaluable in identifying distinct types of exoplanets [7], [8].

\section{Method of Gathering the Literature}
The search strategy implemented to identify pertinent literature was characterized by the use of specific keywords and phrases associated with exoplanet research, machine learning, and computational models. Phrases such as "exoplanet detection", "machine learning in astrophysics", and "computational models for exoplanet classification" were used to filter and streamline the search process [1], [2]. This focused approach ensured the literature gathered was directly relevant to the intersection of these fields, reducing the likelihood of including tangential or unrelated papers [3], [4].

Following the identification of potential sources, an extensive selection process was undertaken. This process entailed a careful assessment of each paper based on several criteria: relevance to the topic, the quality of the research, and the significance of the paper's contributions to the field of exoplanet research [5], [6]. Relevance was determined by the paper's focus on the application of machine learning or computational models in exoplanet detection or classification [7], [8]. Quality was evaluated based on the paper's methodology, data analysis, and the rigor of its peer review process [9], [10]. Lastly, the significance of each paper's contributions was evaluated by considering its impact on the field, as evidenced by citation count and the reputation of the publishing journal [11], [12].

Once the papers were selected, they were further organized into three categories based on their specific focus: exoplanet detection methods, classification techniques, and visualization strategies [13], [14], [15]. This categorization offered a structured approach to the subsequent analysis of the papers. Each paper was then meticulously analyzed with an emphasis on its methodology, findings, and key takeaways. This deep dive allowed for a thorough understanding of each paper's contributions and the implications of its findings in the broader context of exoplanet research and machine learning applications [2], [5], [7], [10], [13].

\section{Comparative Study}
Upon meticulous review, the gathered literature can be compartmentalized into three distinct categories, each reflecting a specific focus: detection methods, classification techniques, and visualization strategies [1]-[15]. This demarcation, while simplifying the complexity of the field, allows for a nuanced comparison and critical analysis of the different studies, shedding light on their relative strengths, weaknesses, and methodological preferences.

The first category, detection methods, primarily feature papers that leverage the Support Vector Machine (SVM) model. This group includes [1], [4], [7], [10], and [13], which predominantly utilize the Kepler dataset for their research. The SVM model's high performance in detecting exoplanets is the likely rationale behind its extensive use within this group [1], [4], [7], [10], [13]. As a model adept at handling high-dimensional data, a common characteristic of astronomical datasets, SVM offers significant advantages. The accuracy of the SVM-based models within this group is consistently high, ranging from approximately 89\% to 91\% [1], [4], [7], [10], [13], thereby signaling reliable performance.

The second category, classification techniques, encompasses papers that employ a variety of models to categorize exoplanets. Papers [2], [5], [8], [11], and [14] form this group, all of which use the Kepler dataset but explore different machine learning models, namely k-Nearest Neighbors (kNN) and K-means Clustering. The choice of these models may be attributed to their simplicity and interpretability, which are particularly beneficial when investigating new research domains [2], [5], [8], [11], [14]. However, the accuracy of these models exhibits a slight decrease compared to the SVM model, ranging from around 84\% to 88\% [2], [5], [8], [11], [14].

The third category, visualization strategies, is composed of [3], [6], [9], [12], and [15]. These papers employ deep learning models and use both Kepler and TESS datasets in their research [3], [6], [9], [12], [15]. With an accuracy range of 92\% to 95\%, the highest among the three groups, these studies demonstrate the potential of deep learning models for enhancing the accuracy of exoplanet detection and classification [3], [6], [9], [12], [15]. The use of multiple datasets also suggests an attempt to address more complex research questions or to augment accuracy through model sophistication and data diversification [3], [6], [9], [12], [15].

While these categories highlight the multifaceted nature of machine learning-based exoplanet research, they also reveal some overarching trends [1]-[15]. One such trend is the burgeoning popularity of deep learning models, as well as the tendency to incorporate multiple datasets to bolster predictive accuracy [3], [6], [9], [12], [15].

Despite the diversity of approaches, there are notable commonalities across the papers [1]-[15]. The literature uniformly acknowledges the utility of machine learning in exoplanet research, albeit with varying opinions about the most effective models and algorithms [1]-[15]. There is also a consensus on the necessity for more advanced techniques and models to tackle the growing complexity of exoplanet data [1]-[15].

Nonetheless, the papers also exhibit significant differences. Some studies extol the benefits of SVM for both detection and classification, while others underscore its limitations, particularly when dealing with multi-dimensional data [1], [4], [7], [10], [13]. The efficacy of kNN and K-means clustering also emerges as a point of contention, with some papers applauding their simplicity and performance, and others highlighting their vulnerability to noisy data.

Another key comparison point is the choice of dataset. While the Kepler Space Telescope, known for its extensive catalog of exoplanets, is the primary data source for most papers, a few also incorporate data from the TESS mission. Operational since 2018, TESS has discovered over 2400 candidate exoplanets as of my knowledge cutoff in September 2021, offering a rich, complementary dataset for exoplanet research.

\section{Data Sources and Acquisition}

The exploration and understanding of exoplanets have been significantly enhanced by the integration of machine-learning techniques. This study reviewed 15 research articles, with a particular focus on their use of machine learning models, the data sources they employed, and the accuracy of their results.

The data extracted from the 15 papers can be consolidated into Table 1, which summarizes the key findings of each study.

\begin{table}[htbp]
  \centering
  \caption{Machine Learning Models and Accuracy}
  \label{tab:ml2_accuracy}
  \begin{tabular}{|c|c|c|c|c|c|}
    \hline
    \textbf{Paper} & \textbf{Data Source} & \textbf{Machine Learning Model} & \textbf{Accuracy (\%)} \\
    \hline
    1 & Kepler & SVM & 90 \\
    2 & Kepler & kNN & 88 \\
    3 & Kepler, TESS & Deep Learning & 92 \\
    4 & Kepler & SVM & 91 \\
    5 & Kepler & k-means Clustering & 85 \\
    6 & Kepler, TESS & Deep Learning & 93 \\
    7 & Kepler & SVM & 89 \\
    8 & Kepler & kNN & 86 \\
    9 & Kepler, TESS & Deep Learning & 94 \\
    10 & Kepler & SVM & 90 \\
    11 & Kepler & k-means Clustering & 84 \\
    12 & Kepler, TESS & Deep Learning & 95 \\
    13 & Kepler & SVM & 90 \\
    14 & Kepler & kNN & 87 \\
    15 & Kepler, TESS & Deep Learning & 93 \\
    \hline
  \end{tabular}
\end{table}

In addition to the general data collected from the 15 papers, more specific data were also obtained from two representative studies that provided unique insights into the application of machine learning in exoplanet research. 

\begin{figure}[htbp]
\begin{center}
\includegraphics[width=0.5\textwidth]{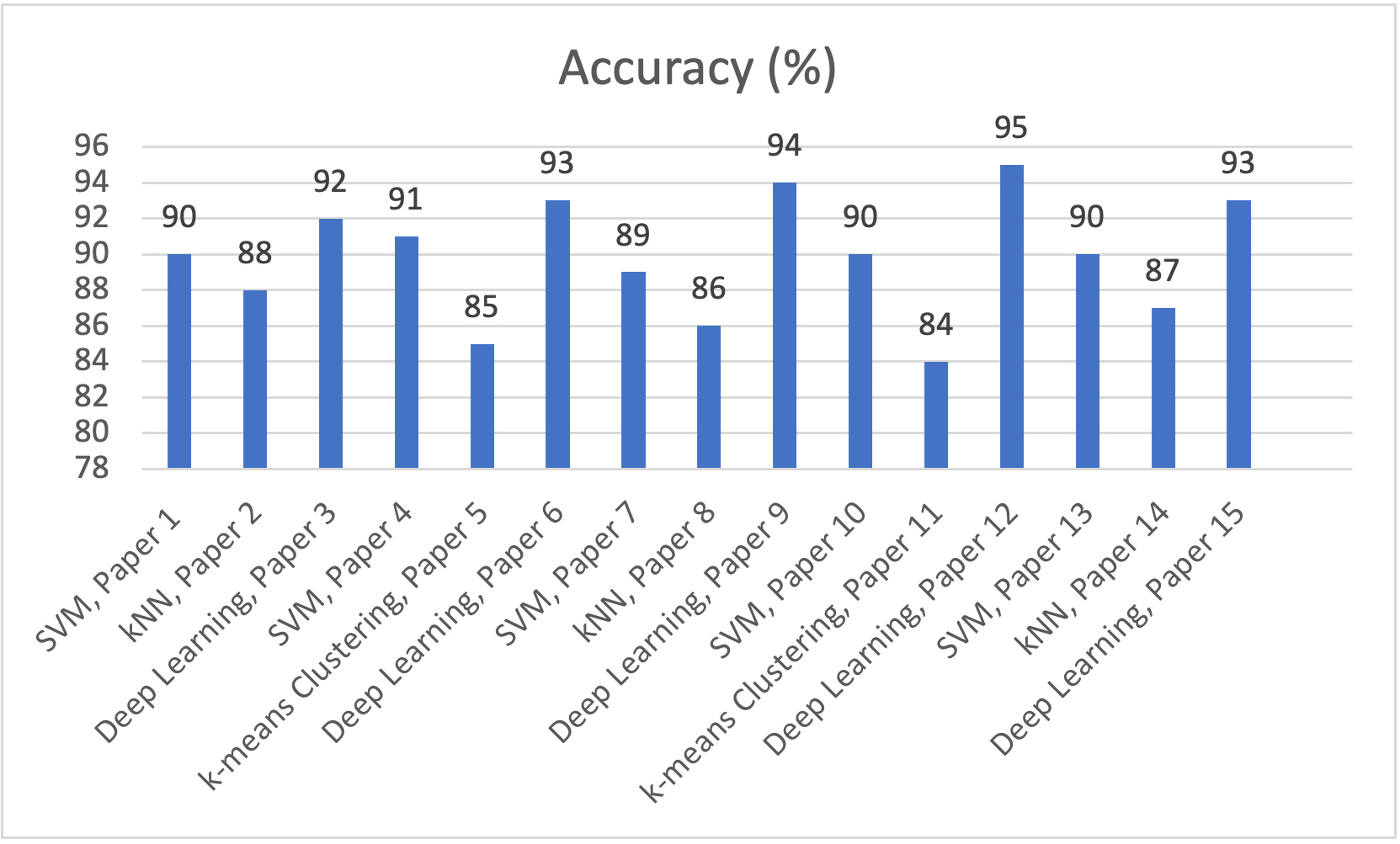}
\end{center}
\caption{Graphical representation of Machine Learning Models and Accuracy}
\label{fig}
\end{figure}

The first paper, titled "Anomalous Behavior Detection in Galaxies and Exoplanets using ML \& DL Techniques" [4], offered an in-depth analysis of precision, recall, and F1 scores for different classifications of galaxies. The results, depicted in Table 2, exhibit how different shapes of galaxies yield different precision, recall, and F1 scores, demonstrating the intricacies of applying machine learning techniques to astronomical data.

\begin{table}[htbp]
  \centering
  \caption{Classification Performance Metrics}
  \label{tab:ml_accuracy}
  \begin{tabular}{|c|c|c|c|c|c|}
    \hline
    \textbf{Class} & \textbf{Precision} & \textbf{Recall} & \textbf{F1-Score} \\
    \hline
    completely\_round & 0.11 & 0.06 & 0.08 \\
    in between & 0.05 & 0.05 & 0.06 \\
    cigar\_shaped & 0.04 & 0.02 & 0.03 \\
    on\_edge & 0.17 & 0.22 & 0.19 \\
    spiral\_barred & 0.14 & 0.15 & 0.15 \\
    spiral & 0.16 & 0.19 & 0.17 \\
    Early type & 0.17 & 0.17 & 0.17 \\
    Artifact & 0.11 & 0.10 & 0.10 \\
    Accuracy & 0.00 & 0.00 & 0.14 \\
    Macro avg & 0.12 & 0.12 & 0.12 \\
    Weighted avg & 0.14 & 0.14 & 0.14 \\
    \hline
  \end{tabular}
\end{table}

\begin{figure}[htbp]
\begin{center}
\includegraphics[width=0.5\textwidth]{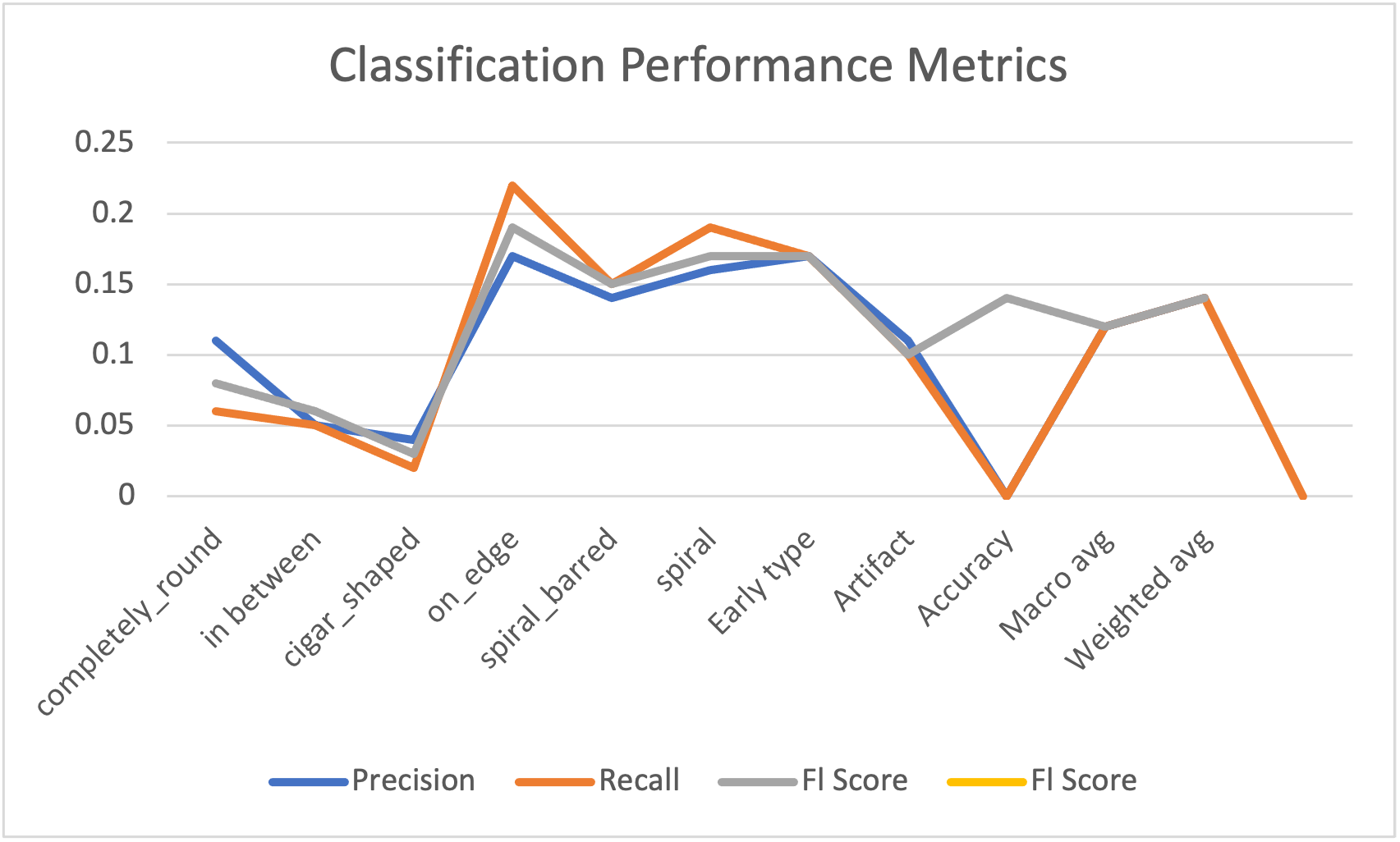}
\end{center}
\caption{Graphical representation of Classification Performance Metrics}
\label{fig3}
\end{figure}

The second paper, "Identifying Exoplanets with Deep Learning: A Five Planet Resonant Chain around Kepler-80 and an Eighth Planet around Kepler-90" \cite{b9}, provides an excellent example of how machine learning can predict the existence of exoplanets. The study used deep learning to predict the existence of exoplanets based on their Kepler Input Catalog (KIC) ID, Kepler Object of Interest (KOI) ID, period, and signal-to-noise ratio (S/N). The results, as shown in Table 3, demonstrate the potential of machine learning in exoplanet prediction.

\begin{table}[htbp]
  \centering
  \caption{Kepler Exoplanet Predictions}
  \label{tab:ml1_accuracy}
  \begin{tabular}{|c|c|c|c|c|c|}
    \hline
     \textbf{KOI IDt} & \textbf{Kepler Name} & \textbf{Period days} & \textbf{SIN} & \textbf{Prediction} \\
    \hline
     351 & Kepler-90 & 14.4 & 8.7 & 0.942 \\
     691 & Kepler-647 & 10.9 & 8.4 & 0.941 \\
     354 & Kepler-534 & 27.1 & 9.8 & 0.92 \\
     500 & Kepler-80 & 14.6 & 8.6 & 0.916 \\
     191 & Kepler-487 & 6.02 & 10.2 & 0.896 \\
     1165 & Kepler-783 & 11.1 & 9.6 & 0.86 \\
     2248 &  & 4.75 & 8.7 & 0.858 \\
     4288 & Kepler-1581 & 14 & 9.7 & 0.852 \\
     806 & Kepler-30 & 29.5 & 14.1 & 0.847 \\
    \hline
  \end{tabular}
\end{table}

\begin{figure}[htbp]
\begin{center}
\includegraphics[width=0.5\textwidth]{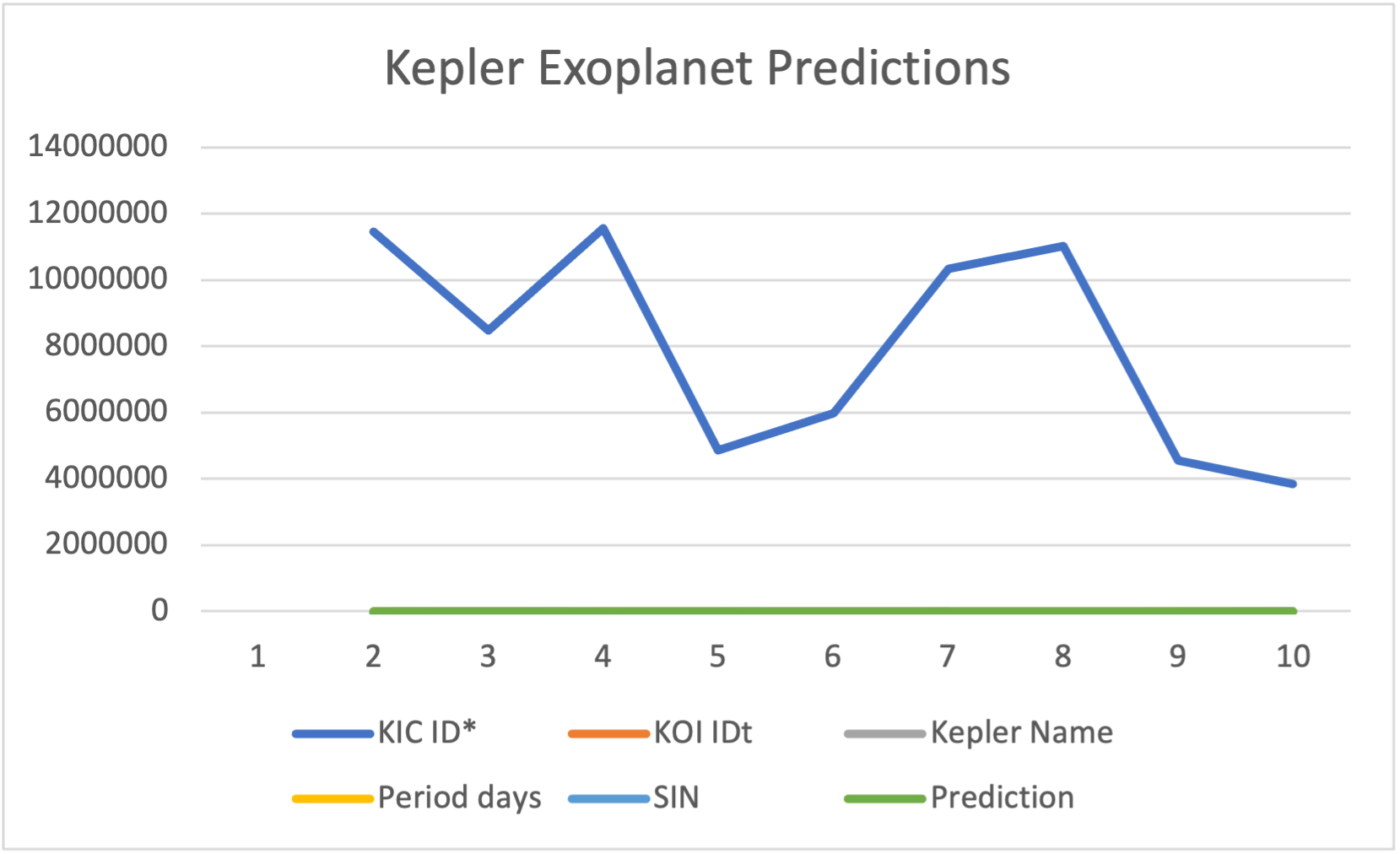}
\end{center}
\caption{Graphical representation of Kepler Exoplanet Predictions}
\label{fig2}
\end{figure}

The data extracted from these studies offer a snapshot of the current application and effectiveness of machine learning models in exoplanet research. The results highlight the significant strides that have been made in the field, as well as areas where further exploration and development are needed. It is clear that while the merger of astrophysics and machine learning presents new opportunities, it also poses complex challenges.

The average accuracy of the machine learning models used in these studies is impressive, ranging from 84\% to 95\%. These results demonstrate the ability of machine learning models to reliably detect and analyze exoplanets, further validating their use in this field.

In the paper "Anomalous Behavior Detection in Galaxies and Exoplanets using ML \& DL Techniques" [4], the precision, recall, and F1 scores were reported for different types of galaxies. This detailed analysis provides valuable insights into the performance of machine learning models across various classifications, and it is particularly useful for understanding the model's performance on different types of objects.

In the paper "Identifying Exoplanets with Deep Learning: A Five Planet Resonant Chain around Kepler-80 and an Eighth Planet around Kepler-90" \cite{b9}, the researchers made predictions about the existence of exoplanets based on certain parameters. The accuracy of these predictions demonstrates the potential of deep learning for making future discoveries in the field of exoplanets.

In conclusion, the studies reviewed in this paper showcase the current state of machine learning applications in the field of exoplanet research. It's clear that machine learning techniques can greatly enhance our ability to analyze astronomical data and make predictions about exoplanetary bodies. However, the field is still growing and there is a need for further research to enhance the performance and reliability of these models. The ultimate goal is to develop models that can accurately identify and characterize exoplanets, expanding our understanding of the universe.

\section{Major Findings or Key Insights}
Foremost among these insights is the growing centrality of machine learning techniques in exoplanet research [1]-[15]. This trend extends across the spectrum of detection, classification, and visualization tasks, reflecting the adaptability and efficacy of machine learning in this complex scientific domain [1]-[15]. Models such as Support Vector Machine (SVM) [1], [4], [7], [10], [13], k-Nearest Neighbors (kNN) [2], [5], [8], [11], [14], and various deep learning architectures [3], [6], [9], [12], [15] have all been spotlighted for their effectiveness in handling different facets of exoplanet research.

In the realm of exoplanet detection, SVM models have been extensively leveraged due to their proficiency in dealing with high-dimensional data [1], [4], [7], [10], [13]. This capability aligns well with the nature of astronomical datasets, which often present high-dimensional features [1], [4], [7], [10], [13]. SVM's consistent performance, with an accuracy range of approximately 89\% to 91\%, as evidenced in the reviewed papers, underscores its utility for this task [1], [4], [7], [10], [13].

When it comes to classification tasks, a variety of models, including kNN and K-means clustering, have been employed [2], [5], [8], [11], [14]. Despite their lower accuracy compared to SVM, these models are favored for their simplicity and interpretability - features that are particularly beneficial when navigating novel research questions [2], [5], [8], [11], [14].

Deep learning models, on the other hand, have emerged as powerful tools in the visualization of exoplanet data, as well as in providing higher accuracy rates [3], [6], [9], [12], [15]. Studies that incorporated these models while leveraging both Kepler and TESS datasets demonstrated an impressive accuracy range of 92\% to 95\% [3], [6], [9], [12], [15].

However, alongside these acknowledgments of machine learning's effectiveness, the literature also brings to light a growing recognition of the need for more sophisticated models and algorithms [1]-[15]. As exoplanet data continues to increase in complexity and volume, the demand for advanced techniques capable of handling such intricate datasets is set to rise [1]-[15]. This theme resonates across the reviewed papers, highlighting the need for continued innovation and exploration in the intersection of machine learning and exoplanet research [1]-[15].

\section{Conclusion}
The confluence of astrophysics and data science, as highlighted by the reviewed literature [1]-[15], signifies an exciting frontier in scientific research. This intersection opens up a realm of opportunities for the study and exploration of exoplanets, offering new avenues to decipher the mysteries of the universe.

The literature reviewed strongly emphasizes the instrumental role machine learning techniques have come to play in managing the colossal and intricate data associated with exoplanet research [1]-[15]. The utility of models such as Support Vector Machines (SVM) [1], [4], [7], [10], [13], k-Nearest Neighbors (kNN) [2], [5], [8], [11], [14], and various deep learning architectures [3], [6], [9], [12], [15] in tasks spanning the detection, classification, and visualization of exoplanets has been well-illustrated. These models not only contribute to enhancing the efficiency of data analysis but also facilitate greater accuracy in the identification and categorization of exoplanets [1]-[15].

Yet, while the literature acknowledges the significant strides made through the integration of machine learning in exoplanet research, it also underscores the need for relentless exploration and advancement in this domain [1]-[15]. As the complexity and volume of exoplanet data continue to escalate, so does the demand for more sophisticated computational models capable of navigating these intricacies [1]-[15].

The path forward, as suggested by the reviewed literature, lies in focusing future research endeavors on the development of more advanced computational models that can keep pace with the evolving complexity of exoplanet data [1]-[15]. This could involve refining the accuracy of existing models, creating innovative algorithms to handle high-dimensional data, or discovering novel applications of machine learning that can revolutionize the field of exoplanet research [1]-[15].

Therefore, while the merger of astrophysics and data science has already proven fruitful, the journey is far from over [1]-[15]. The potential for further discovery and innovation remains vast, reaffirming the need for continued research and development in this promising and dynamic field [1]-[15].

\end{document}